\documentclass[%
reprint,
nofootinbib,
 amsmath,amssymb,
 aps,
]{revtex4-2}

\usepackage{graphicx}% Include figure files
\usepackage{dcolumn}% Align table columns on decimal point
\usepackage{bm}% bold math
\usepackage{multirow}
\usepackage{aas_macros}
\usepackage{color}
\begin{document}

\preprint{APS/123-QED}

\title{Long-term gravitational wave asteroseismology of supernova: \\
from core collapse to 20 seconds postbounce}

\author{Masamitsu Mori}
 \email{masamitsumori@g.ecc.u-tokyo.ac.jp}
\author{Yudai Suwa}%
 \altaffiliation[Also at ]{Center for Gravitational Physics and Quantum Information, Yukawa Institute for Theoretical Physics, Kyoto University, Kyoto
606-8502, Japan.}
\affiliation{%
Department of Earth Science and Astronomy, The University of Tokyo, Tokyo 153-8902, Japan
}%

\author{Tomoya Takiwaki}
\affiliation{
National Astronomical Observatory of Japan, 2-21-1, Osawa, Mitaka, Tokyo, 181-8588, Japan
}%


\date{\today}% It is always \today, today,

\begin{abstract}
We use an asteroseismology method to calculate the frequencies of gravitational waves (GWs) in a long-term core-collapse supernova simulation, with a mass of 9.6 solar mass. The simulation, which includes neutrino radiation transport in general relativity is performed from core-collapse, bounce, explosion and cooling of protoneutron stars (PNSs)  up to 20\,s after the bounce self-consistently. Based on the hydrodynamics background, we calculate eigenmodes of the PNS oscillation through a perturbation analysis on fluid and metric.
We classify the modes by the number of nodes and find that there are several eigenmodes. In the early phase before 1\,s, there are a low-frequency $g$-mode around 0.5\,kHz, a mid-frequency $f$-modes around 1\,kHz, and high-frequency $p$-modes above them. 
Beyond 1 second, the $g$-modes drop too low in frequency and the $p$-modes become too high to be detected by ground-based interferometers. However, the $f$-mode persists at 1 kHz.
We present a novel fitting formula for the ramp-up mode, comprising a mixture of $g$-mode and $f$-mode, using postbounce time as a fitting parameter. Our approach yields improved results for the long-term simulation compared to prior quadratic formulas.
We also fit frequencies using combinations of gravitational mass, $M$, and radius, $R$, of the PNS.
We test three types of fitting variables: compactness $M/R$, surface gravity $M/R^2$, and average density $\sqrt{M/R^3}$. We present results of the time evolution of each mode and the fitting for three different ranges, from 0.2\,s to 1\,s, 4\,s, and 20\,s for each formula. We then compare the deviation of the formulas from the eigenmodes to determine which fitting formula is the best. In conclusion, any combination of $M$ and $R$ fits the eigenmodes well to a similar degree. Comparing 3 variables in detail, the fitting with compactness is slightly the best among them. We also find that the fitting using less than 1\,s of simulation data cannot be extrapolated to the long-term frequency prediction.

%\begin{description}
%\end{description}
\end{abstract}

\maketitle

%\tableofcontents

\section{Introduction}

The gravitational wave (GW) is one of the most important prediction of general relativity, which was directly confirmed by the observation of the binary black hole merger, GW150914 \cite{LIGOScientific:2016aoc}.
After the memorial event, increasing number of GW events have been observed from binary star systems including merger of two neutron stars and black hole-neutron star systems \cite{2021arXiv211103606T}.

The next most promising targets for observation of GWs are supernova explosions, the origin of neutron stars and black holes.
Supernovae (SNe) emit electromagnetic waves, neutrinos, and GWs, and the multimessenger observations can provide deep insight into the supernova interior (see Refs.~\cite{Tony12,Janka12, Kotake12, Burrows13, Foglizzo15, Mueller20,Smartt15, Maeda22} for reviews).
Indeed, the first detection of neutrinos from SN 1987A \cite{Hirata:1987hu,Bionta_1987,Alexeyev_1988} allowed us to estimate the total energy emitted by neutrinos being $\sim 10^{53}$\,erg~\cite{1987PhLB..196..267S, Burrows:1988ba,1989ApJ...340..426L} and led to the conclusion that a neutron star (NS) formed inside supernova explosion and the released gravitational energy drives the explosion. When the next galactic supernova happens, GWs would also be detected~\cite{Abdikamalov_2021}. Combining these independent pieces of information will lead to a breakthrough in the study of supernova explosions.

The GW asteroseismology, which is counterpart of regular light-based asteroseismology  but with GW, has a potential to provide NS parameters \cite{1998MNRAS.299.1059A,2016PhRvD..94d4043S,2017PhRvD..96f3005S,2019PhRvD..99l3024S,2020PhRvD.102f3023S,2020PhRvD.102b3028S,2020PhRvD.102f3025S,2020MNRAS.498.3503S,2021PhRvD.104l3009S,2018MNRAS.474.5272T,2019MNRAS.482.3967T,2019PhRvL.123e1102T,2021PhRvL.127w9901T,2021PhRvD.103f3006B,2019PhRvD.100l3009W}. 
NSs are expected to produce strong GWs from their typical oscillation modes, which is so-called eigenmodes.
If these oscillation modes are observed in GWs, we will be able to extract the combination of the mass $M$ and radius $R$ of the NS. There are three different variations of fitting formulas for evolution of oscillation frequencies; the compactness $(M/R)$, the surface gravity $(M/R^2)$, and the average density $\sqrt{M/R^3}$. For instance, Sotani et al. (2021) \cite{2021PhRvD.104l3009S} proposed a universal relation for models employing different nuclear equations of state based on the average density, while Torres-Forn\'e et al. (2019) \cite{2019PhRvL.123e1102T,2021PhRvL.127w9901T} proposed a comparable formula but using the average density for $p$- and $f$-modes and the surface gravity for $g$-modes. There are also studies where postbounce time is used as a fitting variable. Morozova et al. (2018)~\cite{Morozova:2018glm} used fitting with postbounce time and reported that differences between equation of states showed up. Warren et al. (2020)~\cite{Warren:2019lgb} investigated a correlation between an estimate of GW frequencies and neutrino emission. 

\begin{table*}[htbp]
    \centering
        \caption{Summary of recent GW asterosemismology studies, where $a$ and $b$ of Torres-Forn\'e et al and Mori, Suwa and Takiewaki. are real numbers.}
    \begin{tabular}{cccccc}
    \hline \hline
         Authors & Year & Gravity & Fitting variables & Simulation time (s) & References  \\ \hline
          Sotani and Takiwaki  & 2016 & Newtonian & No fitting & $\sim$1.0 & \cite{2016PhRvD..94d4043S}\\
          Sotani, Kuroda, Takiwaki and Kotake & 2017 & GR & $MR^{-3}$ & $\sim$0.3 & \cite{2017PhRvD..96f3005S} \\
          Morozova et al. & 2018 & Approx GR & Postbounce time & $\sim$1.5 & \cite{Morozova:2018glm} \\
          Torres-Forn\'e et al. & 2019 & Approx GR/ GR & $M^{a}R^{b}$ & $\sim$1.2 & \cite{2019PhRvL.123e1102T,2021PhRvL.127w9901T}\\
          Warren et al. & 2020 & Approx GR & Postbounce time & $\sim$4.0 & \cite{Warren:2019lgb} \\
          Sotani, Takiwaki and Togashi & 2021 & Approx GR & $MR^{-3}$ & $\sim$0.8 & \cite{2021PhRvD.104l3009S}\\
          Sotani and Sumiyoshi & 2021 & GR & $MR^{-3}$ & $\sim$1.4 & \cite{Sotani:2021kvj} \\
          Mori, Suwa and Takiwaki & 202e & GR & $M^{a}R^{b}$, Postbounce time & $\sim$20 & This work \\
          \hline\hline
    \end{tabular}
    \label{tab:summary_prev}
\end{table*}

Table~\ref{tab:summary_prev} summarizes previous studies. 
The number of the models in the general relativistic framework is limited
and no simulations were performed beyond 10\,s except this work.
We have not known how long the fitting formulas are applicable. 
The motivation of this study is to discover long-term behavior of frequencies and find a fitting formula for long-term emission. Another importance of long-term simulation is related to the multimessenger astronomy. If galactic supernovae happen, neutrino events are observable for more than 20\,s~\cite{2019ApJ...881..139S,Mori_2020} so we need the same time prediction of GWs to check the correlation.

Our goal of this work is to discover the behavior of NS eigenmode frequencies and the connection of their properties.
To accomplish this goal, this paper employs the long-term simulation of a supernova explosion and NS formation.
We will utilize data from Ref.~\cite{Mori_2020}, especially long-term (20\,s) self-consistent simulations from the collapse of the iron core to the supernova explosion and the protoneutron star (PNS) cooling phase.
The late period has the great advantage of allowing more precise modeling than the early period ($\lesssim 1$\,s), as the complex physical processes settle down. Recently, Refs.\, \cite{2019ApJ...881..139S,2021PTEP.2021a3E01S,2022ApJ...925...98N,2022ApJ...934...15S} have developed a method to extract the mass and radius of the NS based on theoretical estimates of neutrino emission in the late phase.
Theoretical predictions of GW, on the other hand, have been focused on the early phase, since long-term multi-dimensional numerical simulation are needed to predict the GW.

In this paper, we will show long-term evolution of PNS eigenmodes, which are a source of GW emission, investigate which combination of mass and radius can fit $g$- and $f$- modes the most for a long time and discuss the possibility whether we can estimate late-time frequencies from early time fitting. Our PNS simulation considers the full general relativistic gravity and neutrino transport, and our mode analysis employs metric perturbations. These make our estimation quantitatively precise. Section~\ref{sec:methods} explains the neutrino-radiation hydrodynamics simulation and the method to estimate eigenmodes of GWs based on the simulation. Section~\ref{sec:results} describes the results of the eigenmode analysis and fitting. We use three parameters (compactness, surface gravity, and average density) to fit eigenmodes. We also propose new fitting formula with respect to time after bounce and provide a discussion on which fitting formula is the best. Finally, we summarize our conclusion in Section~\ref{sec:summary}.
Note that we employ the signature $(-, +, +, +)$ as the Minkowski metric and adopt units of $G = c = 1$.

\section{Methods}\label{sec:methods}
This section describes how to estimate the frequency of eigenmodes from our supernova model.
We first conduct core-collapse supernova (CCSN) simulation by solving the neutrino-radiation hydrodynamics equations and then calculate eigenmodes of the PNS based on the simulation results. The calculation of the eigenmodes allows us to compute leading order contribution to GW signals.

\subsection{Neutrino-radiation hydrodynamics simulation}\label{sec:neutrino-radiation-hydro-simulation}

Our CCSN model is based on the model in Ref.~\cite{Mori_2020}.
The progenitor is 9.6$M_\odot$ and zero-initial metallicity (A. Heger, private communication, 2016), which has been reported to explode not only in multidimensional simulations but also in spherical symmetric simulations \cite{Melson_2015, Radice_2017, Mori_2020, 2021PASJ...73..639N}.

As Ref.~\cite{Mori_2020}, we employ a public code, {\tt GR1D} \cite{O_Connor_2010,O_Connor_2015} for our hydrodynamics simulation.
The primitive variables in {\tt GR1D} are density $\rho$, specific internal energy $\epsilon$, velocity $v$ and electron fraction $Y_e$.
The metric used in {\tt GR1D} is 
\begin{equation}
    ds^2 = -\alpha(r, t)^2 dt^2 + X(r,t)^2 dr^2 + r^2d\Omega^2, \label{eq:metric-gr1d}
\end{equation}
where $r$ and $t$ are radius and time. $\alpha$ and $X$ are a lapse and a shift function, which are written with functions of the potential $\Phi$ and enclosed gravitational mass $m(r,t)$,
\begin{align}
    &\alpha = {\rm exp}(\Phi(r, t)), \\
    &X = \sqrt{1 - \frac{2m(r,t)}{r}}.
\end{align}
Those are given as
\begin{align}
    &m(r,t) \\ \nonumber
    &=\pi\int^r_0 (\rho h W^2 - P + \tau^\nu_m)r^{\prime 2} dr^\prime,\\
    &\Phi(r,t) \\ \nonumber
    &=\int^r_0X^2\left[\frac{m(r^\prime,t)}{r^{\prime 2}} +  4\pi r^\prime(\rho h W^2 v^2 + P + \tau^\nu_\Phi)\right]dr^{\prime 2}
    + \Phi_0,
\end{align}
where $h=1+\epsilon+P/\rho$ is enthalpy with $P$ being pressure, $W=1/\sqrt{1-v^2}$ is Lorentz factor with $v=Xv_1$ being the product of three-velocity $v_1$ and $X$ and $\tau^\nu$ is the stress-energy tensor component of neutrinos.
Here, $\Phi_0$ is determined by the matching condition.
The metric must be connected to the Schwarztshild metric at the star's surface, which leads to
\begin{equation}
    \Phi(R_*, t) = {\rm ln}[\alpha(R_*, t)] = \frac{1}{2}{\rm ln}\left[1 - \frac{2m(R_*, t)}{R_*}\right],
\end{equation}
where $R_*$ is the star's radius.

In {\tt GR1D}, the hydrodynamics equations are described as below,
\begin{eqnarray}
    \partial_t \vec{U} + \frac{1}{r^2}\partial_r\left[\frac{\alpha r^2}{X}\vec{F}\right] = \vec{S},
\end{eqnarray}
where $\vec{U}$ is a set of conserved values, $\vec{F}$ is a set of flow values, $\vec{S}$ is a set of source terms, and $\partial_x:=\partial/\partial x$.
To be specific, conserved values are given as
\begin{align}
    \vec{U}=[D, DY_e, S^r, \tau],
\end{align}
where
\begin{align}
    D &= X\rho W,\\
    DY_e &= X\rho W Y_e,\\
    S^r &= \rho h W^2 v, \\
    \tau &= \rho h W^2 - P - D.
\end{align}
The flux vector $\vec{F}$ is
\begin{align}
    \vec{F}=[Dv, DY_e v, S^rv + P, S^r -Dv],
\end{align}
and the source and sink terms are
\begin{align}
    \vec{S} = \left[0, R^\nu_{Y_e}, (S^rv-\tau-D)\alpha X\left(8\pi r P + \frac{m}{r^2}\right)
    + \alpha P X\frac{m}{r^2} \right. \\ \nonumber
    \left. + \frac{2\alpha P}{X r} + Q^{\nu, E}_{S^r} + Q^{\nu, M}_{S^r}, Q^{\nu, E}_{\tau} + Q^{\nu, M}_{\tau}\right],
\end{align}
where $R^\nu_{Y_e}$, $Q^{\nu, E}_{S^r}$, $Q^{\nu, M}_{S^r}$, $Q^{\nu, E}_{\tau}$ and $Q^{\nu, M}_{\tau}$ are the contributions of neutrinos and are calculated through the neutrino transport.

{\tt GR1D} is implemented with the M1 scheme \cite{2011PThPh.125.1255S} with multi-energy groups for neutrino-radiation transport. It solves the Boltzmann equation up to the first two moments and use an analytic closure for closing moment equations.
The energy groups are logarithmically divided into 18 energies. The center value of the lowest energy group is 2\,MeV and that of the highest energy group is 280\,MeV. 
In this simulation, neutrino transport is calculated out to 600\,km and neutrino information is read out at 500\,km considering effects of the gravity.

Interactions between neutrinos and matter are calculated in advance as an opacity table with Nulib. \footnote{https://github.com/evanoconnor/NuLib}
Table~\ref{tab:nu_interations} summarizes interactions used in the simulation. In the original opacity table, the bremsstrahlung is taken into account only for heavy-lepton neutrinos. That is, the interaction of $N + N \rightarrow N + N +\nu_x + \bar{\nu}_x$ is only included. Since we found that this approximation leads to unphysically high average energy of $\nu_e$ and $\bar{\nu}_e$ at the late phase \cite{Mori_2020}, we reproduce the numerical table taking into account the reaction, $N + N \rightarrow N + N +\nu_e + \bar{\nu}_e$.

\begin{table}[htbp]
    \centering
    \caption{Summary of neutrino-matter interactions. Here, $n$ is a neutron, $p$ is a proton, $(A, Z)$ is a nuclei whose mass number is $A$ and atomic number is $Z$. The neutrino interaction with $\nu$ has no sensitivity to flavors but the interaction with $\nu_i$ has a sensitivity to flavors.}
    \begin{tabular}{cc}
        \hline \hline
        Neutrino production & References\\
          \hline
         $\nu_e + n \rightarrow p + e^- $ & \cite{Burrows_2006,Horowitz:2001xf}\\
         $\bar{\nu}_e + p \rightarrow n + e^+$ &  \cite{Burrows_2006}\\
         $\nu_e + (A,Z) \rightarrow (A, Z+1) + e^-$ & \cite{Burrows_2006, Bruenn:1985en}\\
         $e^- + e^+ \rightarrow \nu_x + \bar{\nu}_x$ & \cite{Burrows_2006, Bruenn:1985en} \\
         $N + N \rightarrow N + N +\nu_e + \bar{\nu}_e$ & \cite{Burrows_2006, Bruenn:1985en} \\
         $N + N \rightarrow N + N +\nu_x + \bar{\nu}_x$ &  \cite{Burrows_2006, Bruenn:1985en} \\
         \hline
         Neutrino scattering \\
         \hline
         $\nu + \alpha \rightarrow \nu + \alpha$&\cite{Burrows_2006, Bruenn:1985en} \\
         $\nu_{ i} + p \rightarrow \nu_{ i} + p$&\cite{Burrows_2006, Bruenn:1985en,Horowitz:2001xf}  \\
         $\nu_{ i} + n \rightarrow \nu_{ i} + n$&\cite{Burrows_2006, Bruenn:1985en,Horowitz:2001xf}  \\
         $\nu_{ i} + (A,Z) \rightarrow \nu_{\rm i} + (A,Z)$&\cite{Burrows_2006, Bruenn:1985en, Horowitz_1997}  \\
         $\nu_{i} + e^- \rightarrow \nu^\prime_{i} + e^{-\prime}$&\cite{Bruenn:1985en,1994ApJ...433..250C}\\
         \hline \hline
         
    \end{tabular}
    \label{tab:nu_interations}
\end{table}

\subsection{Asteroseismology}\label{sec:asteroseismology}
In order to calculate eigenmodes of oscillations in PNSs, we employ the asteroseismology method. 
{\tt GREAT}~\cite{2018MNRAS.474.5272T, 2019MNRAS.482.3967T} is open source software for GW asteroseismology 
in general relativity~\cite{2018MNRAS.474.5272T}. That is, the perturbations in the linear analysis both of fluid and metric are considered.
The oscillation follows the next equations:
\begin{align}
   & \partial_r \eta_r +\left[\frac{2}{r} + \frac{1}{\Gamma_1}\frac{\partial_r P}{P}+\frac{\partial_r \psi}{\psi}\right]\eta_r 
   +\frac{\psi^4}{\alpha^2 c_s^2}\left(\sigma^2 - \mathcal{L}^2\right)\eta_\bot \\ \nonumber
    &=  \frac{1}{c^2_s}\frac{\delta \hat{Q}}{Q}-\left(6+\frac{1}{c^2_s}\right)\frac{\delta \hat{\psi}}{\psi},    \label{eq:del_eta_r}\\
    &\partial_r \eta_\bot -\left(1-\frac{\mathcal{N}^2}{\sigma^2}\right)\eta_r + \left[\partial_r {\rm ln}q-\mathcal{G}\left(1+\frac{1}{c^2_s}\right)\right]\eta_\bot 
    \\ \nonumber
    &= \frac{\alpha^2}{\psi^4\sigma^2}\left[\partial_r({\rm ln \rho h})\left(1+\frac{1}{c^2_s}\mathcal{G}\right)\right]\left(\frac{\delta \hat{Q}}{Q}-\frac{\delta \hat{\psi}}{\psi}\right),
    \label{eq:del_eta_bot}
\end{align}
where $\eta_r$ and $\eta_\bot$ are longitudinal and transverse coefficients of eigenmodes respectively, $\sigma$ is the frequency, $c_s$ is the sound speed, $\psi$ is the conformal factor, $Q\equiv \alpha\psi$ and $\Gamma_1\equiv \frac{\rho}{P}\frac{\partial P}{\partial \rho}|_{\rm adiabatic} = h c_s^2$ are the adiabatic index.
$\mathcal{L}^2$ and $\mathcal{N}^2$ are the relativistic Lamb frequency and relativistic Brunt-$\rm V\ddot{a}is\ddot{a}l\ddot{a}$ frequency and their definitions are 
\begin{align}
    \mathcal{L}&\equiv \frac{\alpha^2}{\psi^2}c^2_s\frac{l(l+1)}{r^2}, \\
    \mathcal{N}&\equiv\frac{\alpha^2}{\psi^4}\mathcal{BG},
\end{align}
where $\mathcal{G}$ is the gravity defined as
\begin{align}
    \mathcal{G} =  -\partial_r{\rm ln}\alpha
\end{align}
and $\mathcal{B}$ is the relativistic version of the Schwarzschild discriminant defined as
\begin{align}
    \mathcal{B} = \frac{\partial_r \epsilon}{\rho h} - \frac{1}{\Gamma_1}\frac{\partial_r P}{P}.
\end{align}
Finally, about the metric perturbations, $\delta \hat{\psi}$ and $\delta \hat{Q}$, they read

\begin{align}
   \hat{\nabla}^2\delta\hat{\psi} =& -2\pi\psi^5\left[\left(5\epsilon + \frac{\rho h}{c_s^2}\right)\frac{\delta \hat{\psi}}{\psi} - \frac{\rho h }{c^2_s}\frac{\delta \hat{Q}}{Q}\right] \nonumber\\
    &-2\pi \rho h \psi^5\left(\frac{\psi^5\sigma^2}{\alpha^2 c_s^2}\eta_\bot -\mathcal{B}\eta_r\right),\label{eq:delta_psi}\\
    \hat{\nabla}^2\delta\hat{Q} =& 2\pi(\rho h +5P)\alpha\psi^5\left(\frac{\delta\hat{Q}}{Q} + 4\frac{\delta\hat{\psi}}{\psi}\right)\nonumber\\ 
    +2\pi\rho h\alpha\psi^5
    &\left[\left(6+\frac{1}{c_s^2}\left(\frac{\psi^4\sigma^2}{\alpha^2}\eta_\bot - \frac{\delta\hat{Q}}{Q} + \frac{\delta\hat{\psi}}{\psi}\right)-\eta_r\mathcal{B}\right)\right].
    \label{eq:delta_Q}
\end{align}

In order to find frequencies, $\sigma$, these Eqs.~\eqref{eq:del_eta_r}, \eqref{eq:del_eta_bot}, \eqref{eq:delta_psi} and \eqref{eq:delta_Q} are integrated from the center of the star to the PNS surface ($\rho = 10^{11} {\rm g~cm^{-3}}$). 
The inner boundary condition is
$ \eta_r|_{r=0}=\frac{l}{r}\eta_{\bot}|_{r=0}\propto r^{l-1}$,
and the outer boundary condition is as same as Eq.\,(7) in Ref.~\cite{2020PhRvD.102f3025S}.
See also Eqs.~(4)--(6) of Ref.~\cite{2020PhRvD.102f3025S} to convert the spherical polar coordinate of Eq.~\eqref{eq:metric-gr1d} into the isotropic coordinate used in {\tt GREAT} \cite{2006A&A...445..273M}.

%%%%%%%%%%%%%%%%%%%%%%%%%%%%%%%%%%%%%%%%%%%%%
\section{Results}\label{sec:results}
%%%%%%%%%%%%%%%%%%%%%%%%%%%%%%%%%%%%%%%%%%%%%

%%%%%%%%%%%%%%%%%%%%%%%%%%%%%%%%%%%%%%%%%%%%%
\subsection{PNS properties}
\begin{figure}[htbp]
    \centering
    \includegraphics[width=0.5\textwidth]{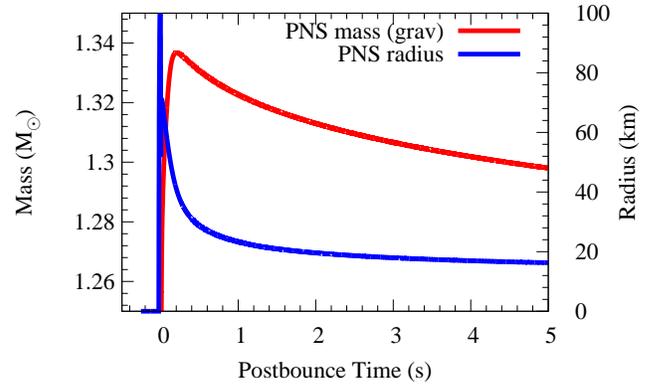}
    \caption{Time evolution of gravitational mass (red) and radius (blue) of the PNS. The left axis indicates the gravitational mass in $M_\odot$ and the right indicates the radius in km. The horizontal axis is the postbounce time in second. }
    \label{fig:NS_mass_radius}
\end{figure}

%%%%%%%%%%%%%%%%%%%%%%%%%%%%%%%%%%%%%%%%%%%%%
Before going to the argument of the GW signal, we briefly give the time evolution of PNS gravitational mass and radius, which are shown in Figure~\ref{fig:NS_mass_radius}.
We define the surface of the PNS at the radius where the density is $10^{11}\,{\rm g\,cm^{-3}}$.
The blue and red lines with the right and the left axis show the radius and gravitational mass of the PNS, respectively.

The PNS radius (blue line) is larger than 100\,km at the bounce and then rapidly shrinks to 13\,km.
The baryonic mass of PNS converges to 1.36\,$M_\odot$ soon after the onset of the explosion. In such a light progenitor, the mass accretion rate is small and PNS mass is converged in the early phase. 
Although the baryonic mass is constant, the gravitational mass (red line) decreases due to the neutrino emission up to 1.26\,$M_\odot$ at 20\,s after the bounce.
Those evolutions are consistent with other studies. For instance, our baryonic mass 1.36$M_\odot$ is consistent with previous studies that used the same progenitor model~\cite{2013ApJ...766...43M, Wanajo:2017cyq, Radice_2017}.
Our gravitational mass at the last moment of the simulation, 1.26$M_\odot$ is also consistent with an approximate estimate given in Ref.~\cite{Radice_2017}.

\begin{figure}[htbp]
    \centering
    \includegraphics[width=0.45\textwidth]{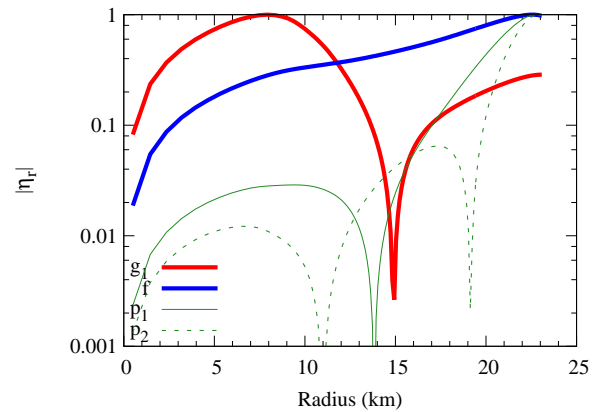}
    \caption{The absolute value of the quantity $\eta_r$ of $g$- (red), $f$- (blue), and $p$-modes (green) in the PNS at 1\,s after the bounce.
The $\eta_r$ is normalized with the maximum values being unity. The solid lines are the eigenmodes which have one node for $g$- and $p$-mode.
The dashed line is the eigenmode which has two nodes.}
\label{fig:amp_f_g}
\end{figure}

\begin{figure*}[htbp]
    \centering
    \includegraphics[width=0.75\textwidth]{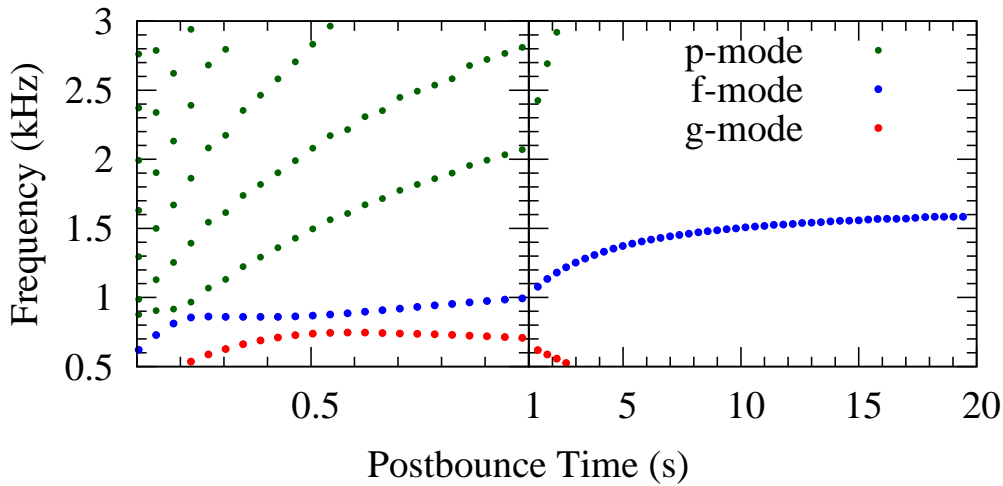}
    \caption{Time evolution of eigenmode frequencies from our supernova simulation. The $g$-mode, $f$-mode and $p$-modes are colored by red, blue and green, respectively. }
    \label{fig:all_modes}
\end{figure*}

%%%%%%%%%%%%%%%%%%%%%%%%%%%%%%%%%%%%%%%%%%%%%
\subsection{Gravitational wave modes}
%%%%%%%%%%%%%%%%%%%%%%%%%%%%%%%%%%%%%%%%%%%%%

Using the model introduced in Section~\ref{sec:neutrino-radiation-hydro-simulation} and the method described in Section~\ref{sec:asteroseismology}, we calculate the oscillation modes of the entire PNS.
In the following, we classify the modes into $p$-modes, $f$-mode and $g$-modes. The modes are characterized by the number of nodes. If there is no node, we call it $f$-mode and otherwise it is $p$- or $g$-modes. The physical difference between $p$-modes and $g$-modes is restoring force. The $p$-modes are invoked by pressure and the $g$-modes are driven by buoyancy. 
The frequency of the $p$-mode is higher as the number of nodes increases. On the other hand, that of the $g$-mode is inversely lower as the number of nodes increases. While this simple classification is also used in Refs.~\cite{2016PhRvD..94d4043S,2017PhRvD..96f3005S,Morozova:2018glm,2018MNRAS.474.5272T,2020PhRvD.102f3023S,2021PhRvD.104l3009S,2020PhRvD.102f3025S},
Refs.~\cite{2019MNRAS.482.3967T,2019PhRvL.123e1102T} employ different classification criteria, but this difference affects only the names of the modes and does not change the following discussion.

Roughly speaking, the $p$-mode propagates near the surface of the star and the $g$-mode propagates near the center of the star.
Figure~\ref{fig:amp_f_g} shows transverse coefficients $\eta_{\rm r}$ of the $f$-mode (blue) and $g$-mode (red) as functions of radius at 1\,s after the bounce, wherein the indexes of the subscript are the number of nodes. The coefficients are normalized for their maximum values to be unity.
There is a node at 15\,km and a peak at 7\,km in the $g_1$-mode. In the $f$-mode, there is no node except at the center and the $\eta_r$ of the $f$-mode increases with the radius. There is a node at 14\,km in the $p_1$-mode and are two nodes at 11\,km and 19\,km. Both $\eta_r$ of the $p$-modes also increases with the radius.

Following the simple mode identification,
we show the time evolution of eigenmodes of GWs in Figure~\ref{fig:all_modes}, which are calculated from time snapshots of our supernova simulation.
We show them from 0.1\,s after the bounce in this figure because matter motion
around the bounce is dynamical and clearly deviated from the eigenmodes.
There are the $g_1$-mode (red), $f$-mode (blue) and $p_i$-modes (green) in Figure~\ref{fig:all_modes}, where $i$ is the natural number and indicates the number of nodes. 

The $g_1$-mode has the lowest frequencies, the $f$-mode is in the middle frequency range, and the $p_i$-modes are the highest. The $g_1$-mode frequencies gradually increase, reach the peak around 0.7\,kHz at 0.55\,s, slowly decrease and eventually pass through 0.5\,kHz at 2\,s.
Such evolution is also seen in the previous studies, e.g., see Fig.~5 of Ref.~\cite{Morozova:2018glm} and
Fig.~3 of Ref.~\cite{2021PhRvD.104l3009S}. The frequencies of $p$-modes increase and even the lowest $p_1$-mode exceeds 3\,kHz at 2\,s. Higher $p_i$-modes increase faster.

The frequencies of the $f$-mode increase from 0.6\,kHz to 0.9\,kHz for the first 0.15\,s, keep the value up to 0.5\,s and then slowly increase again to 1.6 kHz at 20\,s.
There are avoided crossings between the lowest $p$-mode and $f$-mode around 0.2\,s and between the $g$-mode and the $f$-mode around 0.5\,s \cite{Morozova:2018glm,2020MNRAS.498.3503S}.
Note that the term of ``avoided crossing'' means that the frequencies of two eigenmodes approach each other but they do not cross.
In the following section, we focus on the $g$-mode before the avoided crossing and the $f$-mode after the avoided crossing.

\begin{figure*}
    \centering
       \includegraphics[width=0.75\textwidth]{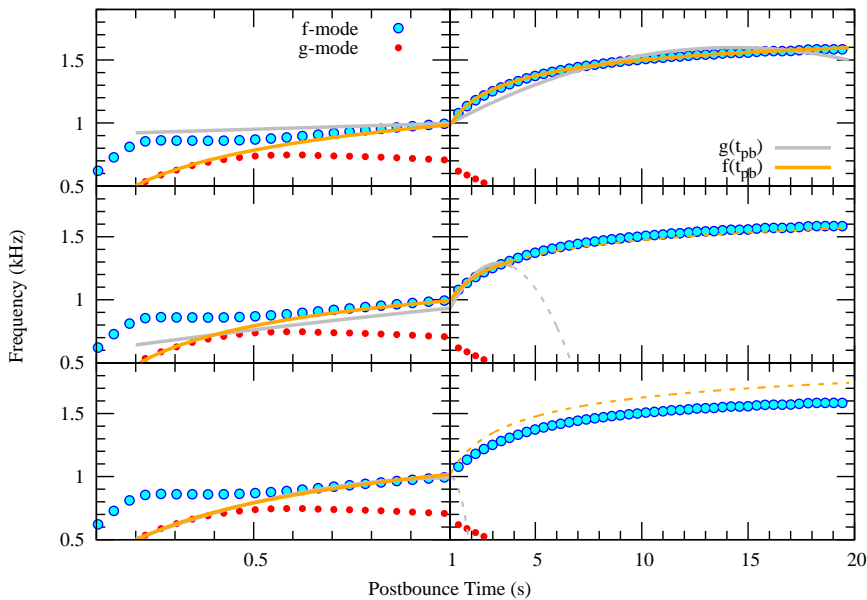}\\
    \caption{Fitting results with postbounce time. 
     The top panel is fitting from 0.2 to 20\,s, the middle is fitting from 0.2 to 4\,s and the bottom panel is fitting from 0.2 to 1\,s. The solid lines are the fitted functions and the dashed lines are extrapolated curves of the fitting. The cyan dots edged with blue are the $f$-mode and the red dots are the $g_1$-mode.
     The orange is the fitting of Eq.~\eqref{eq:g(x)}.
     We also show the fitting results of the quadratic function in gray, $g(x=t_{\rm pb}) = b_1 + b_2x + b_3x^2$ for comparison.}
    \label{fig:fit_time}
\end{figure*}

\subsection{Fitting}\label{sec:fitting}
In this section, we propose new fitting formulas for the eigenmode frequencies that is based on the long-term general relativistic simulation (see Figure~\ref{fig:all_modes}). We provide one fitting formula in terms of postbounce time and three types of fitting methods with respect to the mass and the radius of the PNS: $M/R$, $M/R^2$ and $\sqrt{M/R^3}$, which mean compactness, surface gravity and average density, respectively. See Table~\ref{tab:summary_prev} for the difference from the previous formulas.

We select so-called ramp-up mode that has avoided crossing \cite{Morozova:2018glm,2020MNRAS.498.3503S}, i.e., that is $g$-mode before 0.76\,s and the $f$-mode after 0.76\,s in our classification (see Figure~\ref{fig:all_modes}). In the multi-dimensional simulation, this mode most clearly and ubiquitously appears, e.g., Refs.~\cite{2009ApJ...707.1173M,2013ApJ...766...43M}. In the most of previous studies, fitting of this mode is provided (see Table~\ref{tab:summary_prev}).

First, we use postbounce time as a fitting variable.
Morozova et al.~\cite{Morozova:2018glm} employed quadratic functions to fit eigenmode frequencies with postbounce time. However, the quadratic function can fit curves but cannot fit constant values.
Thus, we also propose a new fitting formula with respect to postbounce time. The function writes:
\begin{align}
    f(x = t_{\rm pb}) = \frac{a_1x^{a_4}}{x^{a_4} + a_2} + a_3,
    \label{eq:g(x)}
\end{align}
where $t_{\rm pb}$ is postbounce time measured in s and $a_1$, $a_2$, $a_3$ and $a_4$ are fitting parameters.
This function is proportional to $x^{a_4}$ when $x$ is close to 0 and becomes constant when $x$ is large enough. 
The parameters determined in this study are shown in Table~\ref{tab:fit_parameters_tpb}. We also fit with quadratic function $(g(x=t_{pb}) = b_1 + b_2x + b_3x^2)$ for comparison with Morozova et al. (2018)~\cite{Morozova:2018glm} and our fitting parameters for the quadratic function are summarized in Table~\ref{tab:fit_parameters_tpb_quadra}.

\begin{table}[htbp]
    \centering
        \caption{Fitting parameters for Eq.~\eqref{eq:g(x)},
    $ f(x) = \frac{a_1x^{a_4}}{x^{a_4} + a_2} + a_3$. Here $x$ is $t_{\rm pb}$. The units of $t_{\rm pb}$ and $g$ are second and kHz, respectively}
    \begin{tabular}{cccccc}
    \hline \hline
        Fitting range (s) & $a_1$ & $a_2$ & $a_3$ & $a_4$ \\
        \hline
0.2-1  & $2.604$ & $0.6971$ & $-0.5158$ & $0.5091$ \\
0.2-4  & $8.488$ & $0.1415$ & $-6.442$  & $0.2907$ \\
0.2-20 & $2.639$ & $0.5371$ & $-0.7313$ & $0.4661$ \\
\hline \hline
    \end{tabular}
    \label{tab:fit_parameters_tpb}
\end{table}

\begin{table}[htbp]
    \centering
        \caption{Fitting parameters for
    $ g(x) = b_1 + b_2x + b_3x^2$. Here $x$ is $t_{\rm pb}$. The units of $t_{\rm pb}$ and $g$ are second and kHz, respectively}
    \begin{tabular}{ccccc}
    \hline \hline
        Fitting range (s)& $b_1$ & $b_2$ & $b_3$ \\
        \hline
0.2-1 & $2.431\times 10^{-1}$ & $1.453\times 10^{0}$ & $-7.043\times 10^{-1}$ \\
0.2-4 &  $5.552\times 10^{-1}$ & $4.471\times 10^{-1}$ & $-6.801\times 10^{-2}$ \\
0.2-20 &  $9.031\times 10^{-1}$ & $9.698\times 10^{-2}$ & $-3.384\times 10^{-3}$ \\
\hline \hline
    \end{tabular}
    \label{tab:fit_parameters_tpb_quadra}
\end{table}

\begin{figure*}
    \centering
    \begin{tabular}{c}
   \includegraphics[width=0.75\textwidth]{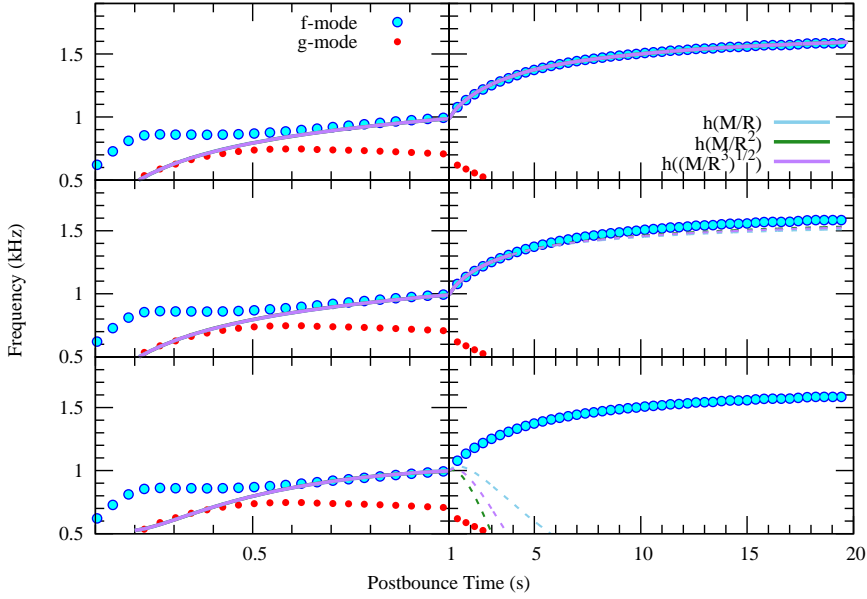}\\
    \end{tabular}
    \caption{Fitting results with mass and radius. The top panel is fitting between 0.2 to 20\,s, the middle is fitting between 0.2 and 4\,s and the bottom panel is fitting from 0.2 to 1\,s. The solid lines are the fitted functions and the dashed lines are extrapolated curves of the fitting. The cyan dots edged with blue are the $f$-mode and the red dots are the $g_1$-mode. The sky-blue is the fitting result of $M/R$, the green is that of $M/R^2$ and the purple is the result of $\sqrt{M/R^3}$.}
    \label{fig:fit}
\end{figure*}

\begin{table*}[]
    \centering
        \caption{Fitting parameters for Eq.~\eqref{eq:f(x)},
    $h(x)= c_1 + c_2\log(x) + c_3x + c_4 x^2$. The units of $M$, $R$ and $h$ are $M_\odot$, km and kHz, respectively.}
    \begin{tabular}{cccccc}
    \hline \hline
        Fitting range (s)& $x$ & $c_1$ & $c_2$ & $c_3$ & $c_4$ \\
        \hline
 \multirow{3}{*}{0.2-1}& $M/R$& $-4.501\times 10^{1}$ & $-9.672\times 10^{0}$ & $4.613\times 10^{2}$  & $-2.459\times 10^{3}$ \\
& $M/R^2$& $-9.209\times 10^{0}$ & $-1.160\times 10^{0}$ & $2.082\times 10^{3}$  & $-3.112\times 10^{5}$ \\
& $\sqrt{M/R^3}$& $-1.836\times 10^{1}$ & $-2.829\times 10^{0}$ & $9.390\times 10^{2}$  & $-3.068\times 10^{4}$ \\
\hline
 \multirow{3}{*}{0.2-4}& $M/R$& $1.048\times 10^{0}$ & $3.222\times 10^{-1}$ & $1.843\times 10^{1}$  & $-5.490\times 10^{1}$ \\
& $M/R^2$& $3.260\times 10^{0}$ & $3.945\times 10^{-1}$ & $4.713\times 10^{1}$  & $-1.497\times 10^{3}$ \\
&$ \sqrt{M/R^3}$& $2.841\times 10^{0}$ & $4.549\times 10^{-1}$ & $2.647\times 10^{1}$  & $-3.361\times 10^{2}$ \\
\hline
 \multirow{3}{*}{0.2-20}& $M/R$& $5.279\times 10^{0}$ & $1.258\times 10^{0}$ & $-1.927\times 10^{1}$  & $1.280\times 10^{2}$ \\
& $M/R^2$& $3.264\times 10^{0}$ & $3.929\times 10^{-1}$ & $3.123\times 10^{1}$  & $1.962\times 10^{3}$ \\
&$ \sqrt{M/R^3}$& $3.340\times 10^{0}$ & $5.303\times 10^{-1}$ & $3.399\times 10^{0}$  & $4.176\times 10^{2}$ \\
\hline \hline
    \end{tabular}
    \label{tab:fit_parameters}
\end{table*}

Figure~\ref{fig:fit_time} shows the fitting results of Eqs.~\eqref{eq:g(x)} (orange) and comparisons to quadratic equations (gray).
We employ data after 0.2\,s in postbounce time because there are no clear eigenmodes due to turbulence around the bounce before this time.
Three panels represent the different fitting ranges of 0.2--1 (bottom), 0.2--4 (middle), and 0.2--20\,s (top).
In the bottom panel, the function $f(x)$ fits the $g$- and $f$-modes well and the result of the quadratic equatiaon overlaps from 0.2 to 1.0\,s.
After 1\,s, which is the extrapolated region, $f(x)$ predicts the higher frequencies. The value is higher by 0.1\,kHz at 10\,s and by 0.15\,kHz at 20\,s. The extrapolation of the quadratic equation does not match the simulation and deviates after 1\,s. As we mentioned above, the quadratic formula is suitable to fit curves but not appropriate for asymptotically constant lines.
The middle panel shows the result of the fitting range spanning from 0.2 to 4\,s. The $f(x)$ matches the simulation overall but predicts a slightly smaller value in the extrapolated region. The difference is 0.02\,kHz at 20\,s. The quadratic function has behavior similar to that in the bottom panel. That is, the function matches before 4\,s but falls down after 4\,s. The way to fall down is slower than that of the fitting from 0.2 to 1\,s.
Finally, the top panel shows the result of the fitting range from 0.2 to 20\,s. The $f(x)$ perfectly matches the simulation and the quadratic function does not match the $g$-mode and has a similar shape after 1\,s.

Next, we fit the eigenmodes with three formulas whose variables are $M/R$, $M/R^2$ and $\sqrt{M/R^3}$, which mean compactness, surface gravity and average density, respectively.
The expression of fitting function is the same as Sotani et al. (2021)~\cite{Sotani:2021kvj}, i.e.,
\begin{align}
    h(x) = c_1 + c_2\log(x) + c_3x + c_4 x^2,
    \label{eq:f(x)}
\end{align}
where $c_1$, $c_2$, $c_3$ and $c_4$ are fitting parameters and the variable $x$ takes $M/R$, $M/R^2$ or $\sqrt{M/R^3}$.

\begin{figure}[htbp]
    \centering
    \includegraphics[width=0.5\textwidth]{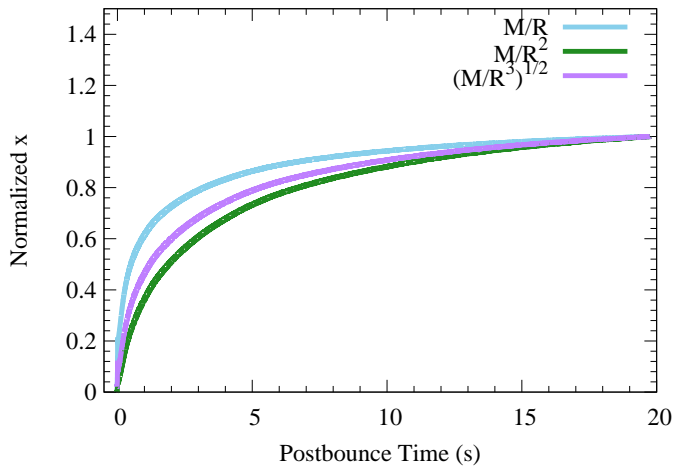}
    \caption{Time evolution of the variables $x=M/R$, $M/R^2$ or $\sqrt{M/R^3}$. All the variables are normalized to be unity at 20\,s.}
    \label{fig:x_t}
\end{figure}

The three variables, $M/R$, $M/R^2$ and $\sqrt{M/R^3}$  behave in the same way.
Figure~\ref{fig:x_t} shows time evolution of the variables.  In the early time, slopes are steep and gradually become flat in the late time. In the early time, the slope of $M/R$ is the steepest, that of $\sqrt{M/R^3}$ is next and that of $M/R^2$ is the most modest.
The normalization is determined with the value at 20\,s.

We fit $h(x)$ of Eq.~\eqref{eq:f(x)} over the different three time ranges: 0.2--1\,s, 0.2--2\,s and 0.2--20\,s in postbounce time as well.
The fitting results are shown in Figure~\ref{fig:fit} and Table~\ref{tab:fit_parameters}. Figure~\ref{fig:fit} shows the fitting lines in the fitting ranges as solid lines and their extrapolations as dashed lines. 
In the case of the fitting range from 0.2 to 1\,s, which is shown in the bottom panel, all the three functions with fitting with $M/R$, $M/R^2$ and $\sqrt{M/R^3}$ are similar and they predict lower frequencies in late time. During the fitting range, they match the simulation well. However, in the extrapolated region, they gradually become lower.
The behavior is the same for all the fitting variables. 
The rate of deviation of $\sqrt{M/R^3}$ is the fastest, followed by $M/R^2$ and finally $M/R$ is the slowest.

In the case of the fitting range from 0.2 to 4\,s, which is shown in the middle panel.
The fitting results of all the variables almost entirely overlap. In the fitting region, they perfectly match the simulation. In the extrapolated region, the fitting functions predict a little smaller values. At 20\,s, the value of the fitting functions is 1.5\,kHz and smaller by 0.09\,kHz than that of the simulation. At last, in the case of fitting from 0.2 to 20\,s, that is, the case that we fit from beginning to end, the all functions reproduce the simulation result well.

\subsection{Comparison with previous studies}
\begin{figure}
    \centering
    \includegraphics[width=0.5\textwidth]{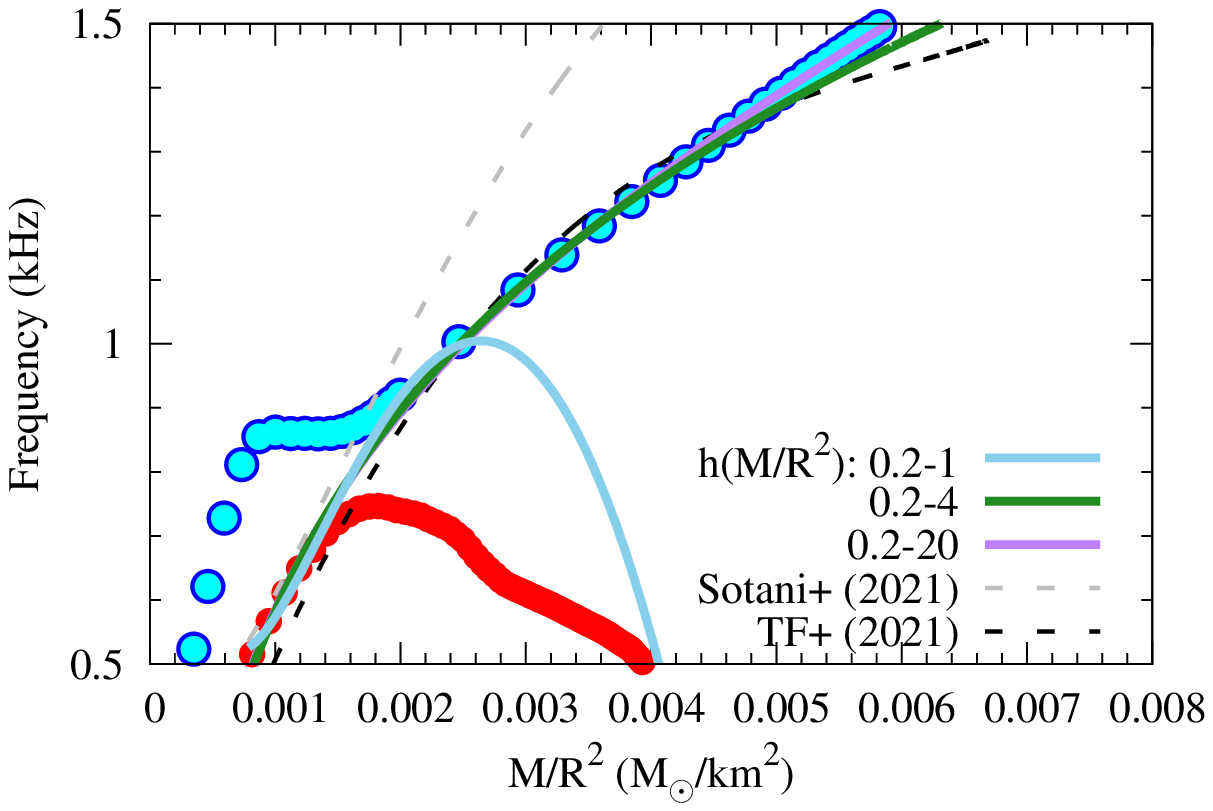}
    \includegraphics[width=0.5\textwidth]{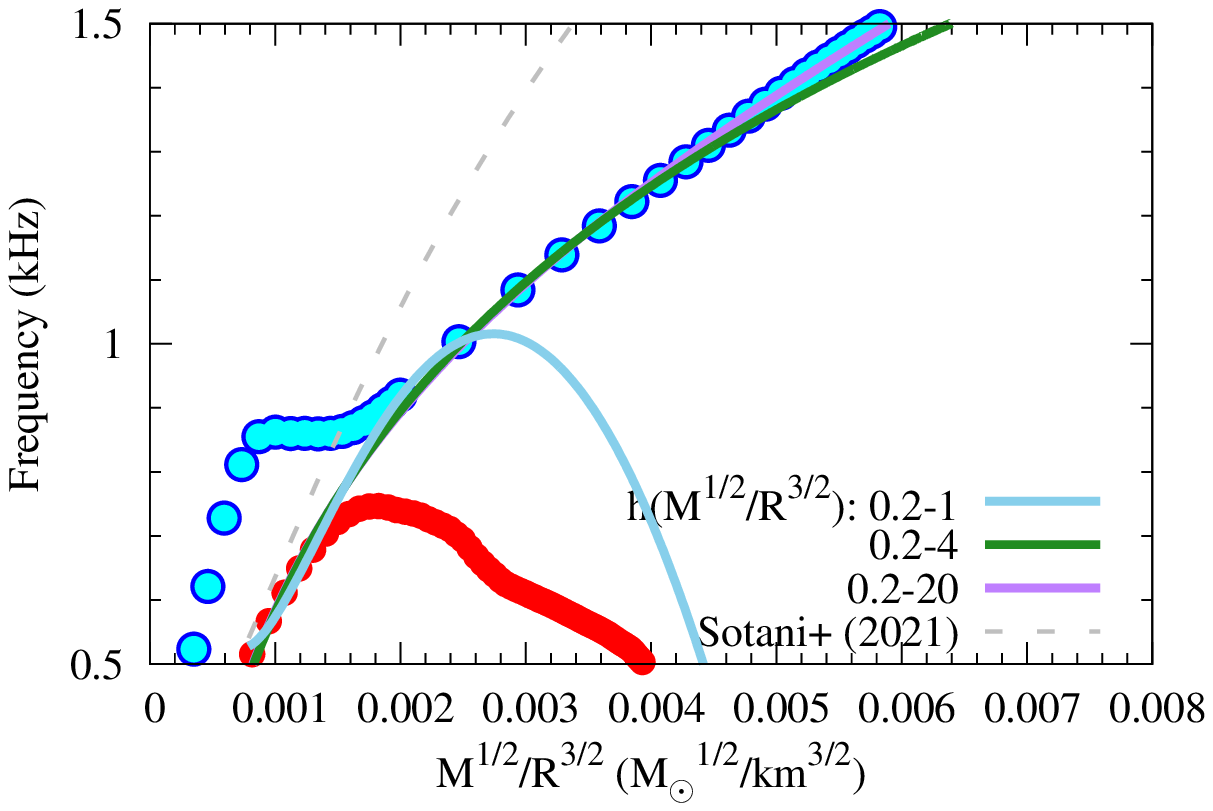}
    \caption{Comparison with fitting results of previous studies. The top panel shows the fitting in terms of $M/R^2$. The gray dashed line is Eq.\,(5) of Sotani et al. (2021)~\cite{Sotani:2021kvj}. The cyan circles edged with blue are the f-mode frequencies and plotted every 0.04\,s before 1\,s and every 0.4\,s after 1\,s. The red circles are g-mode frequencies and plot every 0.04\,s. The black dashed line is the $^2g_{2}$ mode fitting of Torres-Forn\'e et al. (2019)~\cite{2019PhRvL.123e1102T,2021PhRvL.127w9901T}. The bottom panel shows the fitting in terms of $\sqrt{M/R^3}$. The gray line is Eq.\,(3) of Sotani et al. (2021)~\cite{Sotani:2021kvj}.}
    \label{fig:freq_mr2_mr3}
\end{figure}

Figure~\ref{fig:freq_mr2_mr3} compares our fitting formulas with previous studies of Refs. \cite{Sotani:2021kvj, 2019PhRvL.123e1102T}. The horizontal axes are $M/R^2$ in the top panel and $\sqrt{M/R^3}$ in the bottom panel. There are three solid lines of different fitting ranges in each panel. In the top panel, all of the fitting formulas are similar to each other below 1\,kHz. Eq.\,(5) of Sotani et al. (2021)~\cite{Sotani:2021kvj}(gray dashed) leads to slightly higher frequencies overall. The fitting formula of Torres-Forn\'e et al. (2019)~\cite{2019PhRvL.123e1102T,2021PhRvL.127w9901T} overlaps our long-term fitting results. 

In the bottom panel, Eq.\,(3) of Sotani et al. (2021)~\cite{Sotani:2021kvj} (gray dashed) also has higher frequencies than ours. Note that the simulation conditions of our work and previous studies are different, e.g., the progenitor model, treatment of gravity, and equation of state. It is not so strange that the fitting formulas are different as well. For example, Ref.~\cite{2021PhRvD.103f3006B} estimates the error bar of the frequency as $\pm 300$\,Hz using 18 different models (see their Figure 1).

\subsection{Discussion on which fitting is the best}\label{sec:discussion}
According to Figures~\ref{fig:fit_time} and \ref{fig:fit}, the functions in the fitting ranges can reproduce the simulation regardless of variables (except $g(t_{\rm pb})$) and the extrapolation becomes better as the fitting range becomes longer.
This subsection provides a discussion on the fitting results. We here ensure which variable is suitable in detail and how long fitting range is needed.
%\subsection{Deviation of fitting}

We define the dimensionless deviation of fitting below:
\begin{align}
    D(t_{\rm end})&\equiv\int^{T_{\rm sim}}_{T_{\rm start}}\left|\frac{A_{\rm sim}(t) - A_{\rm fit}(x(t), ~t_{\rm end})}{A_{\rm sim}(t)}\right|dt\nonumber\\
    &/\left(T_{\rm sim}-T_{\rm start}\right),
\end{align}
where $t$ is postbounce time, $T_{\rm sim}$ is the simulation time, which is 20\,s now, $T_{\rm start}$ is the starting time of the integral, which is 0.2\,s,  $A_{\rm sim}$ shows eigenmode frequencies as a function of time and $A_{\rm fit}(x(t),~t_{\rm end})$ means fitting functions, $f,g,h$. The fitting range is from 0.2\,s to $t_{\rm end}$ in postbounce time. The smaller value means that the fitting is more accurate.

\begin{figure}
    \centering
    \includegraphics[width=0.5\textwidth]{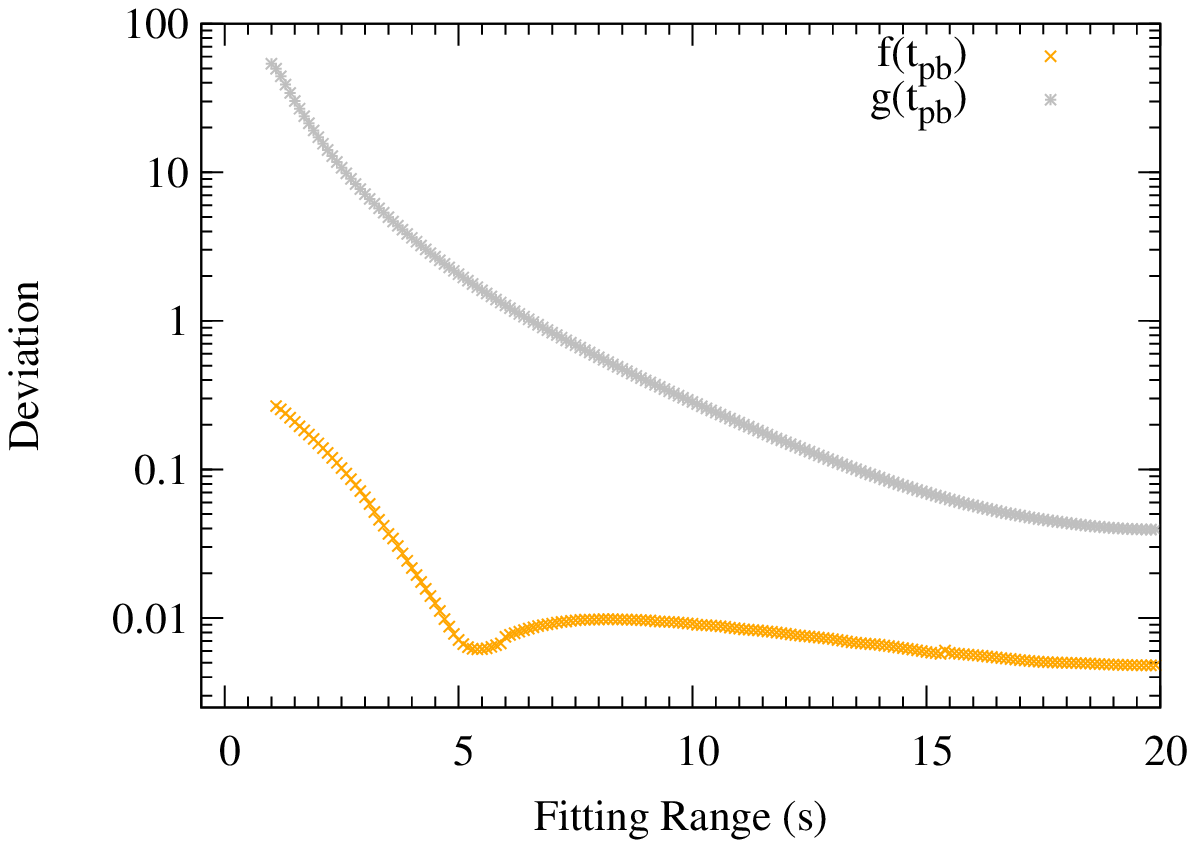}
    \includegraphics[width=0.5\textwidth]{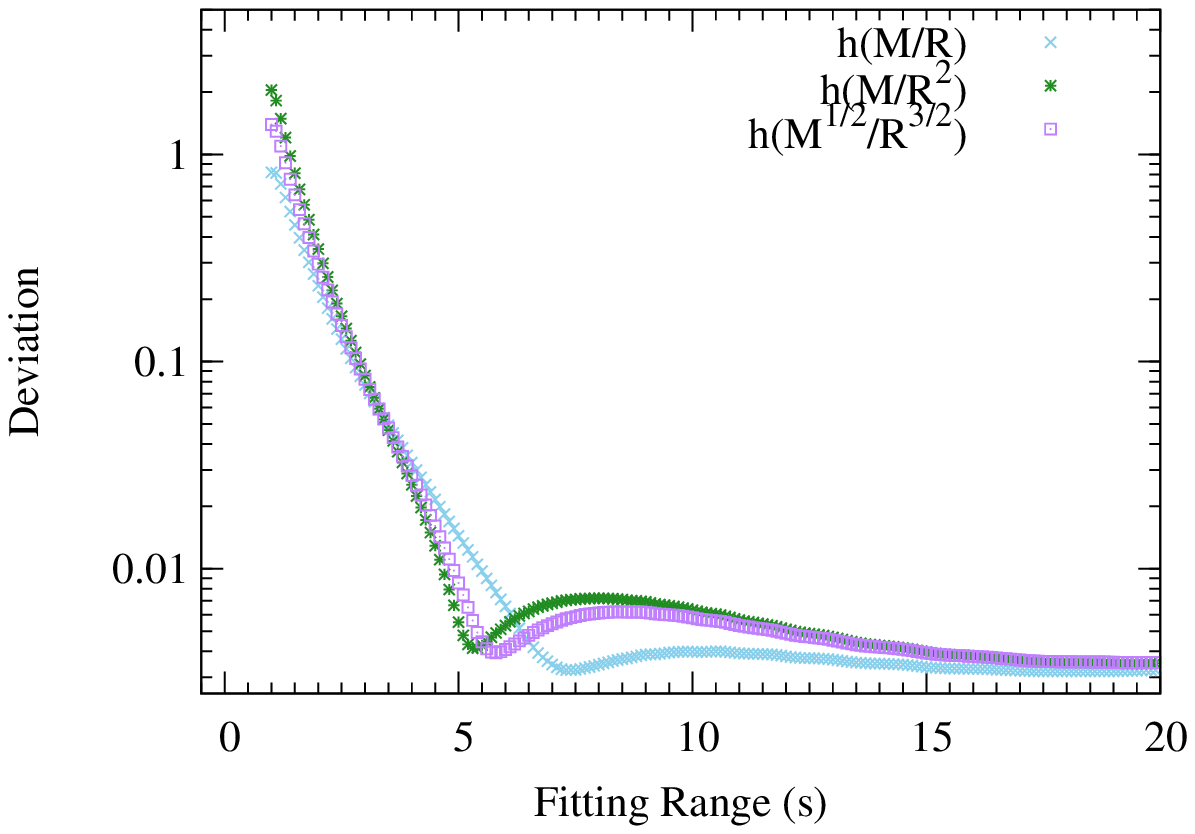}
    \caption{Deviations of each $x$. The horizontal axis is $t_{\rm end}$. The top panel shows the deviations of the functions with respect to $t_{\rm pb}$. The gray is the quadratic function and the orange is Eq.~\eqref{eq:g(x)}. 
    The bottom panel shows those of Eq.~\eqref{eq:f(x)}.
    The sky-blue is $M/R$, the green is $M/R^2$ and the purple is $\sqrt{M/R^3}$.}
    \label{fig:errors}
\end{figure}

Figure~\ref{fig:errors} shows the deviations for functions of each variable from 1\,s to 20\,s in $t_{\rm end}$.
The time bin is 0.005\,s. 
The top panel compares Eq.~\eqref{eq:g(x)} and the quadratic function.
As discussed in the previous section, the quadratic function is not suitable for the long term fitting.
Even if the fitting range is short ($\le$4\,s), $f(t_{\rm pb})$ shows smaller deviation, which is $\sim$ 10\%.
The deviation of $f(t_{\rm pb})$ is about 0.25 at 1.5\,s and gradually decreases to 0.008 at 20\,s.
$f(t_{\rm pb})$ can be used for the rough estimate for the whole evolution of the GW.
The deviation using $f(t_{\rm pb})$ has a local minimum at 5\,s. The fitting slowly converges after 7\,s. A longer fitting range is necessary to obtain the precise estimate ($\le$ 1\%).
Note that, the fitting of $f(t_{\rm pb})$ depends on the initial guess.
 Before 6\,s, we set $(a_1, a_2, a_3, a_4)= (10, 5, 3, 1)$ as the initial guess and $(a_1, a_2, a_3, a_4)= (2.6, 0.5, -0.7, 0.46)$ after 6\,s. On the other hand, the deviation of the quadratic function is much larger. It has 60 at 1\,s and decreases to 0.04 at 20\,s, which corresponds to the deviation of $f(t_{\rm pb})$ at about 3\,s.

The bottom panel shows a comparison of fitting functions of Eq.~\eqref{eq:f(x)} where the fitting parameters are
 $M/R$, $M/R^2$ or $\sqrt{M/R^3}$.
To keep the deviation below percent level, we need simulation for 4\,s at least.
All curves behave in a similar way regardless of fitting variables.
The deviations monotonically decrease until $\sim$ 5\,s ($\sim$ 9\,s for $M/R$) and keep below 0.01 after 5\,s.
Among the variables using $M$ and $R$, the best fit would be $h(M/R)$.
The deviation of fitting with $M/R$ has a maximum of 0.8 at 1\,s, then decreases, has a local minimum of 0.003, raises twice, and finally converges to 0.003. The deviation of $M/R$ is the largest from 4\,s to 6\,s but the deviation is the smallest after 6\,s.
The deviations of fitting with $M/R^2$ and $\sqrt{M/R^3}$ behave similarly. That of $M/R^2$ starts at 2 and has a dip of 0.004 around 5\,s. Then, it has a peak of 0.007 at 7\,s and converges to 0.004 again. Similarly, in the case of $\sqrt{M/R^3}$, the deviation is the biggest value of 1.5 at 1\,s, has a dip of 0.0035
 at 5\,s, has a peak of 0.006 at 7\,s and converges to 0.0035 at last.

\section{Summary}\label{sec:summary}
In this paper, we calculated the frequencies of the eigenmodes of the PNS oscillation based on the long-term SN simulation and 
provided several fitting methods for the GW asteroseismology.
The supernova model was simulated with {\tt GR1D}, which solved the general relativistic neutrino radiation hydrodynamics equations. For the estimate of the frequencies of eigenmodes, we employed {\tt GREAT} that calculates eigenmodes of PNS oscillations.
The calculation continues up to 20\,s, which is the longest compared with recent studies. we fitted the eigenmodes, which are the $g$-mode before the avoided crossing and the $f$-mode after it, with functions considering several types of fitting parameters.

We proposed the new fitting formula using the postbounce time, Eq.~\eqref{eq:g(x)}, and prove that it works better than the simple quadratic function. The quadratic function is suitable for fitting an increasing curve, and it falls short in accurately depicting an asymptotically constant one.

We also derived fitting equations using $M/R$, $M/R^2$, and $\sqrt{M/R^3}$ as the previous studies \cite{2017PhRvD..96f3005S,2019PhRvL.123e1102T,2021PhRvL.127w9901T,2021PhRvD.104l3009S,Sotani:2021kvj}.
These formulas effectively fit the eigenmodes, however, using $M/R$ leads to a slightly better long-term fit compared to the other two variables.
Nevertheless, the difference is small, not making it a matter of choice between the variables.
We also found that the fitting using less than 1\,s of simulation data cannot be extrapolated to the long-term frequency prediction.

In order to give the quantitative behavior of the GW emission, we need to conduct multi-dimensional simulations~\cite{2015ApJ...811...86Y,Kuroda:2017trn,2017MNRAS.468.2032A,OConnor:2018tuw,2019ApJ...876L...9R,2020PhRvD.102b3027M,Nakamura:2022zlc,Andersen:2021vzo,Bugli:2022mlq,Bruel:2023iye}. The multi-dimensional simulation costs too many computational resources.
However, Figure~\ref{fig:errors} indicates that simulations up to 5\,s are enough for giving reliable GW predictions.

For future observations, our goal is to estimate properties of the supernova through the GWs. For this, we have to know inverse functions of eigenmode and $x$ of Eq.~\eqref{eq:f(x)}.  Especially, in the late time, the turbulence of fluid subsides. If we use the late-time information, we can estimate the mass and the radius of the neutron star.
In future work, we will prepare a lot of systematic simulations in order to make a template of eigenmodes and make extrapolations of radii and masses of neutron stars.
The long-term template of eingenmodes and fluid properties allow us to quickly extract information on supernova interior.
Moreover, in the case of a galactic supernova, we can also observe supernova neutrinos, which allow us to do multi-messenger astronomy. From a point of view of multi-messenger astronomy, it is worth estimating properties of SN and PNS independently with different messengers such as GWs and neutrinos.
Indeed, there is a method to estimate PNS masses and radii from supernova neutrinos~\cite{2022ApJ...934...15S,2022MNRAS.512.2806N,Nakazato:2020ogl}. By combining neutrinos and GWs, we can check the consistency and give more reliable estimates than independent analysis. 
Since our supernova simulation includes the neutrino radiation transport, the combined analysis is possible, which will be reported in the future.

\section*{Acknowledgement}
This work is supported by JSPS KAKENHI (JP18H01212, JP20H00174, JP20H01904, JP21H01088, JP22H01223) and Grant-in-Aid for Scientific Research on Innovative Areas (JP18H05437, JP20H04747, JP22H04571) from the Ministry of Education, Culture, Sports, Science and Technology (MEXT), Japan.
This work was supported by computer clusters in Center for Computational Astrophysics.
This research was also supported by MEXT as ‘Program for Promoting researches on the Supercomputer Fugaku’ (towards a unified view of the universe: from large-scale structures to planets, JPMXP1020200109) and JICFuS.

%america oak ridge(mezzacappa, bruenn), princeton (burrows)
%german janka
%austlaia bernhard muller
%spain cerda-duran, Obergaulinger
%japan kuroda

%\nocite{*}

\bibliography{bib}% Produces the bibliography via BibTeX.

\end{document}